\documentclass[prb,twocolumn,showpacs]{revtex4}
\usepackage{graphicx}
\usepackage{amsmath}
\usepackage{amsxtra}
\usepackage{amssymb}
\usepackage{dcolumn}
\usepackage{float}
\usepackage{bm}
\usepackage[breaklinks=true,colorlinks,citecolor=blue,linkcolor=blue,urlcolor=blue]{hyperref}

\renewcommand{\Re}{\mathrm{Re}\,}
\renewcommand{\Im}{\mathrm{Im}\,}

\renewcommand{\Re}{\mathrm{Re}\,}
\renewcommand{\Im}{\mathrm{Im}\,}

\DeclareMathAlphabet{\bi}{OML}{cmm}{b}{it}
\def\be{\begin{equation}}
\def\ee{\end{equation}}
\def\bearr{\begin{eqnarray}}
\def\eearr{\end{eqnarray}}
\def\la{\langle}
\def\ra{\rangle}

\begin{document}
\title{Wave packet dynamics in monolayer MoS$_2$ with and 
without a magnetic field}

\author{Ashutosh Singh}
\affiliation{Department of Physics, Indian Institute of Technology-Kanpur, Kanpur 208016, India}
\author{Tutul Biswas}
\affiliation{Department of Physics, Indian Institute of Technology-Kanpur, Kanpur 208016, India}
\author{Tarun Kanti Ghosh}
\affiliation{Department of Physics, Indian Institute of Technology-Kanpur, Kanpur 208016, India}
\author{Amit Agarwal}
\email{amitag@iitk.ac.in}
\affiliation{Department of Physics, Indian Institute of Technology-Kanpur, Kanpur 208016, India}

\date{\today}

\begin{abstract}
We study the dynamics of electrons in monolayer Molybdenum Disulfide (MoS$_2$), in the absence as well as presence of a transverse magnetic field. 
Considering the initial electronic wave function to be a Gaussian wave packet, 
we calculate the time dependent expectation 
value of position and velocity operators. In the absence of the magnetic field, the 
time dependent average values of position and velocity show damped oscillations dependent on the width of the wave packet. In the presence of a transverse magnetic field, the wave packet amplitude shows oscillatory behaviour over short timescales associated with classical cyclotron orbit, followed by the phenomena of spontaneous collapse and revival over larger timescales. We relate the timescales of these effects and our results can be useful for the interpretation of experiments with trapped ions.

\end{abstract}

\maketitle
 
\section{INTRODUCTION}

In recent years, the interest in atomically thin two-dimensional materials with 
honeycomb lattice structure have grown considerably after the
realization of single layer graphene \cite{grphn1,grphn2}. 
Apart from graphene there are other crystals like silicene
\cite{sili1,sili2}, Boron Nitride\cite{hbn}, 
monolayer group-VI dichalcogenides {\mbox MX}$_2$ 
(\mbox{M=Mo, W} and \mbox{X=S, Se}) etc. which 
also have a honeycomb lattice structure.
Currently, the dichalcogenide material MoS$_2$
(Molybdenum disulphide) \cite{mos2} has drawn lot of attention due to its 
unique electronic\cite{elec}, optical \cite{opt,opt1} and 
transport \cite{trans} properties. 
Like graphene, the Brillouin zone of monolayer {\mbox MoS}$_2$ consists of two 
inequivalent ${\bf K}$ points. 
However, unlike graphene {\mbox MoS}$_2$ monolayer is a direct band-gap 
semiconductor with a gap\cite{bandgp1} of the order of $1.6$ eV.
Since the band gap lies in the visible frequency range, it is also suitable for optoelectronic device applications.

In 1930, Schr\"{o}dinger \cite{Schro} pointed out that a relativistic 
electron in vacuum, described by the Dirac equation, exhibits rapid 
oscillations, which are manifested in the time evolution of physical observables such as 
position, current and spin angular momentum. This phenomenon, also known 
as {\it zitterbewegung} (ZB),  occurs due to the superposition\cite{huang} 
of electron and hole counterparts of the Dirac spinor. 
Experimental observation of ZB has been illusive till date due to
the large oscillation frequency $ \omega_{\rm z} \simeq 10^{21} $ Hz and 
the small oscillation amplitude $ \lambda_{\rm z} \simeq 10^{-13} $ m. 
In addition,  it was shown by Lock \cite{lock} that the ZB oscillations of 
Dirac electrons described by a wave packet decay with time.

{\it Zitterbewegung} is not necessarily a relativistic effect. It can also
occur in non-relativistic electrons, {\it e.g.}, in crystalline solids 
\cite{solid,solid1}, and is a direct consequence of coupling between the energy eigenstates of the system.
In 2005, Zawadzki \cite{zbS1} showed that the oscillation amplitude 
of ZB is enhanced by a factor of $10^4$ in narrow gap semiconductors.
In the same year, Schliemann et. al. \cite{spinzb1} studied the wave packet 
dynamics and ZB in spin-orbit coupled two-dimensional electron gas 
formed at the heterojunctions. These works triggered a large number of 
theoretical investigations of wave packet dynamics and ZB in various 
condensed matter systems \cite{cms1,cms2,cms3, zbgen1,zbgen2, tutul}, including 
carbon nanotubes \cite{cnt}, graphene 
\cite{zbgraph,zbgrph1,zbgrph2,zbgrph3,zbgrph5,Schliemann, Romera_PRB, Kramer, Peters1}, superconductors \cite{zbgen1} and 
hole Luttinger systems \cite{zblutt}.
The wave packet dynamics and ZB have also been studied in 
other systems such as 2D photonic crystal\cite{zbopp}, 
2D sonic crystal \cite{zbsonic} and spin-orbit coupled 
atomic gases \cite{zbbec1}.
The relation between amplitude of ZB oscillations and 
Berry connection was explicitly shown in Refs.~[\onlinecite{zbbec1,berry1,berry2}].
Recently experimental observations of ZB phenomenon 
has been reported in trapped ion systems \cite{Gerritsma} as well as in $^{\bf 87}$Rb Bose Einstein condensates \cite{zbrb1,zbrb2}.

In this Article we study the dynamics of electrons described 
by an initial Gaussian wave packet in monolayer MoS$_2$. 
In the absence of an external magnetic field, we find the expected behaviour of oscillations and decay 
of the wave packet amplitude, which is manifested in the 
the expectation values of physical observables such as position and velocity. 
In the presence of a perpendicular magnetic field, {\it i.e.}, for a discrete Landau level spectrum, 
the long term dynamics of the wave packet shows the phenomena of spontaneous collapse and revival \cite{Perelman} of the wave packet amplitude, which is a direct example of the so called {\it quantum recurrence theorem} \cite{QRT}.

This Article is organized as follows.
In section II we discuss low energy Hamiltonian,  the electronic eigenstates and eigen-energies 
of a MoS$_2$ monolayer. We then study the dynamics of physical observables such as the position and velocity, 
and find that these are characterized by  particular oscillation frequencies and decay constant. 
In section III we explore the dynamics of a Gaussian wave packet in the presence of a transverse magnetic field. 
Starting with the Landau level spectrum and the corresponding wave functions 
of MoS$_2$ monolayer, we calculate the long term time evolution of physical observables of an initial Gaussian wave packet. 
We find the phenomena of spontaneous collapse and consequent revival here, for which the associated timescales follow
from the expansion of the wave packet energy around some dominant Landau level \cite{Perelman, Nauenberg}. 
We summarize our work in section IV.

\section{Dynamics of a Gaussian wave packet}
\subsection{Hamiltonian}

The single particle low-energy Hamiltonian of monolayer MoS$_2$, in the vicinity of each of the two valley's 
at the edges of the hexagonal first Brillouin zone (BZ), denoted by ${\bf K}$ and ${\bf K^{\prime}}$ respectively, 
is given by \cite{Xiao, Rose} 
\begin{eqnarray}\label{Ham}
H=\big\{\hbar v_F(\zeta q_x\tau_x+q_y\tau_y)+\Delta_0\tau_z\big\} \otimes \openone_{\sigma}+H_{\rm so}~,
\end{eqnarray}
where $v_F$ is the Fermi velocity, ${\bf q}=(q_x,q_y)$ is the wave vector of 
electron around ${\bf K}$ points, and $\Delta_0$ is the `direct' band gap at the corner points of hexagonal BZ.
Here $\zeta = +1$ denotes the valley ${\bf K}$ and $\zeta = -1$ denotes the ${\bf K^{\prime}}$ valley. 
The set of Pauli  matrices given by $\tau_i$ and $\sigma_i$ act on the conduction band-valance band space 
and the electron-spin space, respectively, 
and the symbol $\otimes$ denotes the direct product.
In Eq.~\eqref{Ham} the term $H_{\rm so}$ represents the Hamiltonian 
corresponding to the spin-orbit interaction,  which can be described by the following expression
\be
H_{\rm so}=\zeta\Delta_{\rm so}^c\frac{\openone_\tau+\tau_z}{2}\otimes \sigma_z+ 
\zeta\Delta_{\rm so}^v\frac{\openone_\tau-\tau_z}{2}\otimes \sigma_z~,
\ee
where 
$2 \Delta_{\rm so}^c( 2\Delta_{so}^v)$ is the spin-orbit gap in the 
conduction (valence) band. Typically in MoS$_2$, 
$\Delta_{\rm so}^c\approx3$ meV and $\Delta_{\rm so}^v\approx150$ meV so that the gap $\Delta_{\rm so}^c$
can be neglected and the effective SOI gap can be approximated as $\Delta_{\rm so}\approx\Delta_{\rm so}^v$. 
We note that even though $H_{\rm so}$ is different in the two valleys,  it is a constant for 
each valley, and the system is equivalent to two spin-resolved Dirac Hamiltonians  given by, 
\be \label {haml_tot}
H=\hbar v_F(\zeta q_x\tau_x+q_y\tau_y)+\Delta_{\zeta s}\tau_z+ \frac{\Delta_{\rm so}^c+\Delta_{\rm so}^v}{2} \openone_\tau~,
\ee
where the spin and valley dependent gap is given by 
$\Delta_{\zeta s}=\Delta_0-\zeta  s \Delta_{\rm so}/2$ with $s =\pm 1$ denoting the 
up-spin and down-spin electron respectively. The last term of constant energy,  
$(\Delta_{\rm so}^c+\Delta_{\rm so}^v)/2 $, plays no physical role and will be neglected henceforth.
We emphasise here that Eq.~\eqref{haml_tot}, is a spin and valley resolved Hamiltonian 
whose low energy spectrum coincides with the massive Dirac equation in two dimensions - with a gap that is dependent on the 
spin and valley index.

The energy eigenvalues of the spin and valley resolved Hamiltonian, in Eq.~\eqref{haml_tot}, are given by 
\begin{eqnarray}
 \epsilon_{\bf q}^\lambda=\lambda
 \sqrt{(\hbar v_Fq)^2+\Delta_{\zeta s}^2}~,
\end{eqnarray}
where $\lambda = \pm 1$ denotes the conduction and valance bands respectively. 
The corresponding eigen-functions are given by
\begin{eqnarray}
\psi_{\bf q}^\lambda({\bf r})=\frac{e^{i{\bf q}\cdot{\bf r}}}{2 \sqrt{2}\pi}
\begin{pmatrix}
 1\\ \lambda e^{-i\theta}
\end{pmatrix},
\end{eqnarray}
where $\tan{\theta}=q_y/q_x$.

In the rest of the Article, we will `inject' an electronic Gaussian wave function in monolayer MoS$_2$, and study its dynamics. A generic wave packet for MoS$_2$ will have a eight-component spinor form, but since the spin and valley parts of the low energy Hamiltonian are decoupled from each other, each of the two-component spinors (corresponding to conduction and valance bands) evolve independently of each other. Henceforth we only consider the evolution of `spin and valley polarized' two-component spinor.
\subsection{Time evolution of physical observables}
To describe the time evolution of electron 
wave packets we start with an electronic Gaussian wave packet ({\it i.e.}, a spinor wave packet having 
the conduction band component as unity and no valance band component), 
\begin{eqnarray}
\Psi({\bf r},0) = \frac{1}{2\pi}\int d^2q ~
a_{\bf q}~e^{i{\bf q}.{\bf r}}\begin{pmatrix} 1\\0\end{pmatrix},
\end{eqnarray}
where $a_{\bf q} =
\frac{d}{\sqrt{\pi}}e^{-\frac{d^2}{2}(|{\bf q}-{\bf q_0}|)^2}$ is a
Gaussian wave packet of width $d$, which is initially centered 
at ${\bf q}={\bf q_0}$, in the momentum space. Note that $d$ should be greater than the lattice spacing, 
so that $\Psi$ is a smooth enveloping function. 
The wave packet at a
later time $t$ can be found by operating the time evolution operator, $U(t)=
e^{\frac{-iHt}{\hbar}} $, on the initial state {\it i.e.}, $\Psi({\bf r},t) = U(t)\Psi({\bf r},0)$.
Straightforward  calculation gives the spinor wave packet at a later time $t$ to be of the following  form in the Fourier space:

\begin{equation}\label{wave_t}
\begin{pmatrix}\Phi_1({\bf q},t)\\
 \Phi_2({\bf q},t)
\end{pmatrix}
 =  \begin{pmatrix} 
 \cos(\Omega_{\bf q} t)-i\frac{\Delta_{\zeta s} }{\hbar \Omega_{\bf q}}
 \sin(\Omega_{\bf q} t)
\\ -i\frac{ v_F}{\Omega_{\bf q}} (\zeta q_x+iq_y) \sin(\Omega_{\bf q} t)
\end{pmatrix}, 
\end{equation}
where $\hbar \Omega_{\bf q} \equiv \sqrt{\Delta_{\zeta s}^2+(\hbar v_F q)^2}$ is half of the 
difference between the two energy states of the conduction and valance bands, 
{\it i.e.}, $\epsilon_{\bf q}^+$
and $\epsilon_{\bf q}^-$, respectively. The real space spinor components, 
are now simply given by   
$[\Psi_1 ({\bf r},t), \Psi_2 ({\bf r},t)] = 
\int d^2q ~e^{i{\bf q} \cdot {\bf r}} ~[\Phi_1 ({\bf q},t), \Phi_2 ({\bf q},t)]$.

The expectation value of position operator can now be calculated using the 
momentum representation of the position operator: ${\bf \hat{r}_{\rm op}}=i\nabla_{\bf q}$, and is 
given by the following equation,
\be
\langle {\bf r}(t) \rangle = i\int d^2q\, \Phi^{\dagger}({\bf q},t)
\nabla_{\bf q} \Phi({\bf q},t)~.
\ee
Straightforward calculation yields
\begin{eqnarray}
\langle {\bf r}(t)\rangle&=&\langle {\bf r}(0)\rangle + \frac{\Delta_{\zeta s} v_F^2}{\hbar} \int
d^2q \frac{|a_{\bf q}|^2}{\Omega_{\bf q}^2} 
\Big(t-\frac{1}{2\Omega_{\bf q}}\sin(2{\Omega_{\bf q}}t)\Big) \bf{q}\nonumber\\
&-&\zeta \frac{v_F^2}{2}\int d^2 q ~
\frac{|a_{\bf q}|^2 q^2}{\Omega_{\bf q}^2}~{\nabla_{{\bf q}}
\theta}\Big\{1-\cos(2\Omega_{\bf q}t)\Big\}~.
\end{eqnarray}
For the specific choice of $a_{\bf q}$ adopted here, it is easy to show that 
$\langle {\bf r}(0)\rangle=0$. For definiteness, in the rest of the manuscript we choose, $\zeta = 1$ and $s=1$, and define $\Delta \equiv \Delta_{\zeta s}$.
Now, by considering the initial momentum of the 
Gaussian wave packet along $y$ direction ({\it i.e.}, $q_{0x}=0$ and $q_{0y}=q_0$)
one can obtain the expectation value of $x$  component of the position operator (`{\it transverse'} to the initial momentum of the wave packet) to be

\be \label{posx}
\langle x(t)\rangle= \int_0^\infty d\tilde{q} \eta_{\tilde q} \left[1-\cos\Big(2\frac{v_Ft}{d}\tilde{\Omega}_{\bf q}\Big)\right]~,
\ee
where we have defined 
\be \label{eq:eta}
\eta_{\tilde q} = de^{-q_0^2d^2} \frac{{\tilde{q}}^2}{\tilde{\Omega}_{\bf q}^2}
e^{-{\tilde{q}}^2}I_1(2q_0d\tilde{q})
\ee
and  $I_1(x)$ is the (first order) modified Bessel function of second kind and
we have defined the dimensionless parameters: $\tilde{q}=qd$, 
$\tilde{\Delta}=(d/\hbar v_F)\Delta$ and 
$\tilde{\Omega}_{\bf q}
=(d/ v_F)\Omega_{\bf q}$. 
Note that for $x \gg1$, the modified Bessel function of the second kind has 
the following asymptotic form, $I_1(x) = e^{x}/\sqrt{2 \pi x}$.

\begin{figure}[t]
\begin{center} 
\includegraphics[width=1.0 \linewidth]{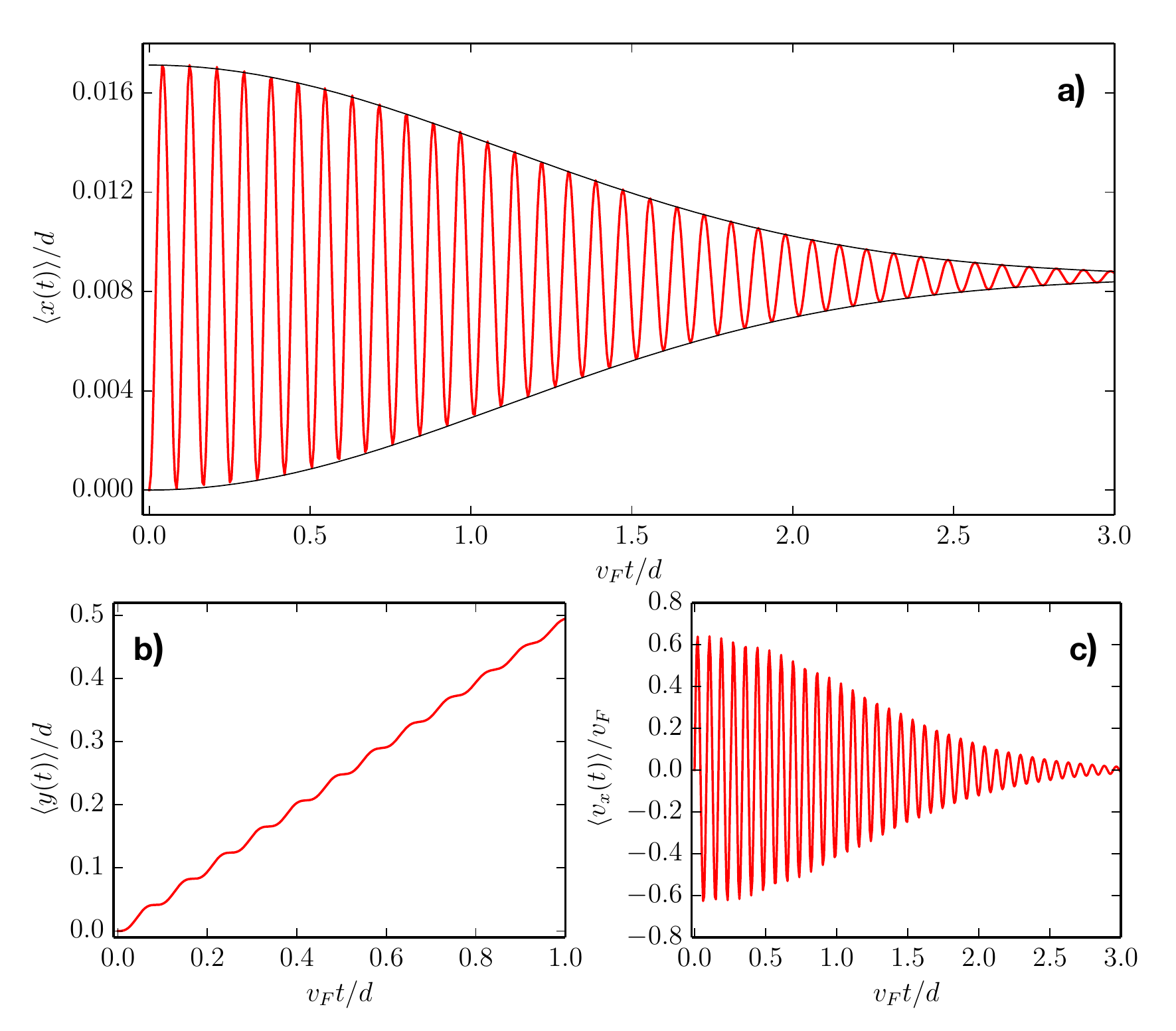}
\end{center}
\caption{Panel a) shows `transverse' {\it zitterbewegung}, {\it i.e.}, expectation value of $ x $ versus $t$. The solid black line highlights the decaying envelop of the oscillating position amplitude, which is given by Eq.~\eqref{eq:decay}.
In Panel b) we plot the `longitudinal'  {\it zitterbewegung} or $\la y \ra$ versus $t$. In panel c) 
we study the expectation value of $ v_x $ versus $t$. 
The wave packet amplitude oscillations in all the panels occur over a timescale $\tau_{\rm zb}$ given by Eq.~\eqref{eq:tauzb}, and the decay in the amplitude occurs over a 
timescale $\tau_{\rm d}$, given by Eq.~\eqref{eq:decay}.
We have chosen the monolayer MoS$_2$ parameters $\Delta = 0.8$ eV, $v_F = 85000$ m/s, with $d = 2\times 10^{-9}$ m, and $q_0 d = 24$ for these plots.
}
\label{fig1}
\end{figure}

The integrations over $\tilde{q}$ in 
Eq.~\eqref{posx} is performed numerically,  
and the time dependence of the position expectation 
value is shown in Fig.~\ref{fig1}a).  Two prominent features observed in Fig.~\ref{fig1}a) are,  
the oscillations or ZB in the expectation value and  
the decay of the oscillation amplitude of ZB. The oscillation timescale of the ZB 
can be extracted, by working in the $q_0 d \gg 1$ limit where one can use the asymptotic form 
of the modified Bessel function given above, and noting that in Eq.~\eqref{eq:eta}, $\eta_{\tilde{q}} \approx  d e^{-(\tilde q - q_0 d)^2} \tilde{q}^2/(\tilde{\Omega}_{\bf q}^2 \sqrt{4 \pi q_0 d \tilde{q}})$. This implies that the integrand in Eq.~\eqref{posx} is peaked around $q_0d$ and consequently the oscillation timescale of ZB is given by 
\be \label{eq:tauzb}
\tau_{\rm zb}= \frac{\pi d}{ v_F\sqrt{\tilde{\Delta}^2 + q_0^2 d^2}}~.
\ee 
The decay in the oscillation amplitude of ZB, occurs due to the interference of different frequencies, corresponding to different $\tilde{q}$ in Eq.~\eqref{posx}.
This decay timescale can be obtained from Eq.~\eqref{posx}, by using the exponential form of $\cos(2 v_Ft \tilde{\Omega}_{\bf q}/d) = \Re [e^{i (2 v_Ft \tilde{\Omega}_{\bf q}/d) }$] to include the time dependence in the exponential terms. Straightforward algebra gives the 
decay of the oscillation amplitude to be of the form: $e^{-t^2/\tau_{\rm d}^2}~$, where  
\be \label{eq:decay}
\tau_{\rm d}^{-2} =  \frac{ v_F^2}{d^2} \frac{q_0^2 d^2}{q_0^2 d^2 + \tilde{\Delta}^2}~.
\ee
As a check of our estimate we note that Eq.~\eqref{eq:decay} fits very well to the exact numerical results --- see the solid black lines in Fig.~\ref{fig1}a).

The $y$ component of the position operator (`{\it longitudinal'} to the initial momentum of the wave packet) is given by 
\be \label{posy}
\langle y(t)\rangle = 2 \tilde{\Delta} \int_0^\infty d\tilde{q} \eta_{\tilde q} 
\left[\frac{v_Ft}{d}-\frac{1}{2\tilde{\Omega}_{\bf q}}\sin\Big(2\frac{v_Ft}{d}\tilde{\Omega}_{\bf q}\Big)\right]~,
\ee
where $\eta_{\tilde q}$ was defined in Eq.~\eqref{eq:eta}.  
As expected due to finite initial momentum in the $y$ direction, the centre of the wave packet moves in the $y$ direction with a constant velocity. 
The ZB oscillations occur around the centre of the moving wave packet on the timescale given by $\tau_{\rm zb}$, but with reduced amplitude,
due to the presence of $\tilde{\Omega}_{\bm q}$ in the denominator of the oscillating term --- see Fig. \ref{fig1}b). Eventually, for $t \gg \tau_{\rm zb}$, the oscillating contribution to 
 $\la y \ra$ becomes vanishingly small in comparison to the term linearly increasing with time.

The components of the velocity operator can be obtained from the Heisenberg equation of motion: 
$ i \hbar v_i=[x_i,H]$. The expectation value of the velocity components can now be calculated by using the spinor 
wave functions in Eq.~\eqref{wave_t}, and are given by 
\be \label{velcx}
\langle v_x(t)\rangle=\frac{2v_F}{d} \int_0^\infty d\tilde{q} \eta_{\tilde q} \sin\left(2 \frac{v_Ft}{d}\tilde{\Omega}_{\bf q}\right)~,
\ee
and
\be \label{velcy}
\langle v_y(t)\rangle=\frac{2v_F}{d} \tilde{\Delta} \langle x(t)\rangle~ .
\ee
As a check of our calculations, we note that $\la v_x(t) \ra$ and $\la v_y(t) \ra$ 
can also be obtained directly by differentiating Eq.~\eqref{posx} and Eq.~\eqref{posy}
with respect to time. In Fig.~\ref{fig1}c). we plot the time dependence of the $x$ component of the velocity operator 
and it shows a behaviour similar to that of the position operator.

We note in passing that in the $\Delta \to 0$ limit, the MoS$_2$ energy spectrum reduces to graphene 
energy spectrum. In the same limit, our result reproduce the known results for wave packet dynamics 
in monolayer graphene \cite{zbgrph3} (in the absence of  magnetic field).  In particular,  Eq.~\eqref{posx} of our manuscript reduces to Eq.~(17b) of Ref. [\onlinecite{zbgrph3}]. 

\section{Wave packet dynamics in a perpendicular magnetic field: spontaneous collapse and revival}

In this section we study the cyclotron dynamics of an electron wave packet in a magnetic field 
${\bf B} = (0,0,B)$, perpendicular to the  monolayer MoS$_2$ plane.  Before we calculate the long term dynamics of the injected wave packet, it is useful to review the timescales associated with the dynamics in systems with discrete energy levels.

\subsection{Oscillation, cyclotron and revival timescales}
Any 2D electronic system subjected to a transverse magnetic 
field, leads to  the formation of discretized Landau levels. The dynamical evolution of wave packets of a quantum system with discrete but non-equidistant 
energy spectrum is generally quiet complex due to quantum interference. However several well defined periodicities emerge \cite{Perelman, Nauenberg, Robinett, Romera_PRB, Demikhovskii_PRA}, if the initial wave packet is
peaked around some large Landau level characterized by $n_0$.

Various timescales which emerge for a discrete systems with non-equidistant energy spectrum,  can be understood from the approximate analytic form of the autocorrelation function\cite{Nauenberg} of the wave packet, which is defined as $A(t) = \langle \Psi({\bf{r}},t) |\Psi({\bf{r}},0) \rangle$. Expanding this wave packet in terms of the orthonormal eigenstates of the system under consideration, $\psi_n$, labelled by discrete Landau level index $n$, we get $ \Psi({\bf{r}},t) = \sum_n c_n \psi_n({\bf r}) e^{- i \epsilon_n  t/\hbar}$, where $\epsilon_n$ are the discrete energy eigenvalues of the system and $c_n = \langle \psi_n ({\bf{r}})| \Psi({\bf{r}}, 0)\rangle$. The autocorrelation function is now given by  
 \be
 A(t) =  \sum_n |c_n|^2 e^{ i \epsilon_n  t/\hbar}~.
\ee 
For studying localized wave packets which are centred around a large  value $n=n_0$, with spread $\delta n$, such that $n_0 \gg \delta n  \gg 1$, we can approximate the form of the expansion coefficients to be a Gaussian distribution centred around $n_0$ with spread of $\delta n$. 
For such cases it is appropriate to do a Taylor series expansion of the energy, $\epsilon_n  = \epsilon_{n_0} + (n-n_0) \epsilon_{n_0}' + (n-n_0)^2 \epsilon_{n_0}''/2 + \dots $, where $\epsilon_n' = (d\epsilon_n/dn)_{n=n_0} $ and so forth. The autocorrelation function is now given by 
\be
A(t) = \sum_{n = - \infty}^{\infty} |c_n|^2 e^{i t/\hbar \left( \epsilon_{n_0} + (n-n_0) \epsilon_{n_0}' + (n-n_0)^2 \epsilon_{n_0}''/2 + \dots \right) }~,
\ee
and each term in the exponential defines a characteristic timescale via, 
\be \label{timescale1}
\tau_{\rm osc} = \frac{2 \pi \hbar}{\epsilon_{n_0}} ~,~~~ \tau_{\rm cl} = \frac{2 \pi \hbar}{|\epsilon_{n_0}'|}~, ~{\rm and}~~~ \tau_{\rm rev} = \frac{4 \pi \hbar}{|\epsilon_{n_0}''|}~.
\ee
The timescale $\tau_{\rm osc}$, is an intrinsic `{\it zitterbewegung}' time scale, which leads to a $n$-independent overall phase which induces {\it no interference} in the wave packet and is thus unimportant for studying long term dynamics of the system. At the `classical' cyclotron time $\tau_{\rm cl}$, the wave packet comes back to the initial position and the autocorrelation function approximately reaches its initial value. 
For the `quantum' revival of the wave packet the terms proportional to the second derivative in the energy should be in multiples of $2 \pi$, and this leads to additional recurrences of the initial wave packet at time $ t = \tau_{\rm rev}$. This revival hierarchy continues to higher orders. Moreover, at times that are rational fractions of $t_{\rm rev}$,  the wave packet undergoes fractional revivals, {\it i.e.}, a sequence of reconstructions, which give  regular well-localised structure of the wave packet amplitude. The fractional revival at $\tau_{\rm rev}/2$ is particularly interesting, as it represents the initial wave packet shifted by half of the classical period \cite{Perelman} ({\it i.e.}, $\tau_{\rm cl}/2$).

In addition to these, for $ \tau_{\rm cl} \ll t < \tau_{\rm rev}$, the terms quadratic in $n-n_0$ lead  to the dephasing and consequently collapse of the initial wave packet over timescale given by,  $\tau_{\rm coll} = \tau_{\rm rev}/(\delta n)^2$. 
The exact expressions for these timescales for the present case of monolayer MoS$_2$ are given in the next subsection.
See Ref.~[\onlinecite{Robinett}] for a detailed review regarding these timescales.

\subsection{Evolution of a Gaussian wave packet}
We now proceed to calculate the time evolution of the initial wave packet. 
To obtain the Landau level spectrum of MoS$_2$ in a transverse magnetic field, we work with 
Landau gauge ${\bf A}=(-By,0,0)$ for the vector potential. Now  making the Landau-Peierls substitution  
${\bf q} \to {\bf q}+e{\bf A}/\hbar$ in Eq.~\eqref{haml_tot}, 
the spin and valley resolved single-particle Hamiltonian is given by 
\be \label{Ham_mag}
H_{\bf B} = \hbar v_{F} \left[\zeta \left(q_x-\frac{eB}{\hbar} y\right) \tau_x + q_y \tau_y \right] + \Delta \tau_z~.
\ee
The energy spectrum of the Hamiltonian in Eq.~\eqref{Ham_mag}, which forms quantized Landau levels, 
is given by \cite{Rose}
\be \label{Enr}
\epsilon_n^\lambda = \lambda\sqrt{\Delta^2 + \varepsilon^2n}~,
\ee
where $\varepsilon=\sqrt{2} \hbar v_F/l_c$ with $l_c=\sqrt{\hbar/eB}$ being  the magnetic length and $\lambda = \pm 1$.
The corresponding eigen-vectors for $\lambda = \pm 1$ branches are  given by
\begin{eqnarray}\label{wavfnP}
\psi_{n,q_x}^+(x,y) = \frac{e^{iq_xx}}{\sqrt{2\pi A_n}}\begin{pmatrix} 
-D_n\phi_{n-1}(y-y_c)\\ \phi_n(y-y_c)\end{pmatrix} ~,
\end{eqnarray}
\begin{eqnarray}\label{wavfnM}
 \psi_{n,q_x}^-(x,y) = \frac{e^{iq_xx}}{\sqrt{2\pi A_n}}\begin{pmatrix} \phi_{n-1}(y-y_c)
 \\D_n\phi_n(y-y_c)\end{pmatrix},
\end{eqnarray}
where 
\begin{eqnarray}
\phi_n(y-y_c) &=& N_n~e^{-(y-y_c)^2/2l_c^2}~H_n[(y-y_c)/l_c]~, \nonumber \\
D_n &=& (\Delta + \sqrt{\Delta^2 +\epsilon^2n})/(\epsilon\sqrt{n})~, \nonumber \\
A_n& =& 1+D_n^2~, 
\end{eqnarray}
 with 
$N_n=(\sqrt{\pi}~2^n~n!~l_c)^{-1/2}$, $H_n$ is the Hermite polynomial of order $n$,
and finally $y_c=q_xl_c^2$. 
Note that for $n=0$ there is only one state with energy
\be
\epsilon_0=\Delta~,
\ee
and eigen function
\be
\psi_0(x,y)= \frac{e^{iq_xx}}{\sqrt{2\pi}}\begin{pmatrix} 0\\\phi_0(y-y_c)
\end{pmatrix}~.
\ee

We now calculate the expectation values of 
position and velocity operator in presence of a finite transverse 
magnetic field following the Green's function approach\cite{zbgrph3}.  
Since $q_y$ is no longer a good quantum number of the system, it is simpler to 
work in the position representation for the initial wave packet (unlike the $B=0$ case where it was 
easier to work with the Fourier representation of the initial wave packet). 

We consider 
the initial state to be a coherent state in a magnetic field, {\it i.e.}, a Gaussian wave packet of the following form,
\begin{eqnarray}\label{ini_mag}
\Psi({\bf r},0)=\frac{1}{\sqrt{\pi}l_c}\exp\Big({-\frac{r^2}{2l_c^2}+i q_{0}x} \Big)~
\begin{pmatrix}
 1\\0
\end{pmatrix}~,
\end{eqnarray}
where $q_{0}$ is the initial momentum along the $x$ direction and  the width of the 
Gaussian wave packet is considered to be equal to the magnetic length, $l_c$. 
The wave packet at a later time $t$ can be written as 
\begin{eqnarray}\label{wavp_t}
\Psi({\bf{r}},t)=\int G({\bf{r}},{\bf{r^\prime}},t)
\Psi({\bf{r^\prime}},0)d{\bf{r^\prime}},
\end{eqnarray}
where $G({\bf r},{\bf r^\prime},t)$ is the Green's function which 
is a $2\times2$ matrix in this case.
The matrix elements of the Green's functions are given by
\begin{eqnarray}
G_{\mu\,\nu}({\bf{r}},{\bf{r^\prime}},t) = \sum_{n,\lambda} 
\int d{q_x}\psi_{n,q_x,\mu}^{\lambda}({\bf{r}},t)
{\psi_{n,q_x,\nu}^{\lambda^\ast}}({\bf{r^\prime}},0),
\end{eqnarray}
where $\psi_{n,q_x}^\lambda$'s are given by Eqs.~\eqref{wavfnP}-\eqref{wavfnM} and 
$\psi_{n,q_x}^\lambda({\bf r},t)=\psi_{n,q_x}^\lambda({\bf r},0)e^{-i\epsilon_n^\lambda t/\hbar}$
with $\epsilon_n^\lambda$ being the energy eigenvalues given in Eq.~\eqref{Enr}.
Since we are injecting the initial wave packet in the $\lambda =+1$ branch only, {\it i.e.}, populating only the positive 
Landau level energies, [see Eq.~\eqref{ini_mag}], 
only two elements $G_{11}$ and $G_{21}$ of the Green's function matrix are required to find 
the components of the wave packet at a later time $t$. Now $G_{11}$ and $G_{21}$ are 
given by
\begin{eqnarray}\label{Grn11}
 G_{11}({\bf r},{\bf r}^\prime,t) &=& \frac{1}{2\pi}\int_{-\infty}^{+\infty}dq_xe^{iq_x(x-x^\prime)}\nonumber\\
 &\times&\sum_{n=0}^{\infty}P_{n+1}\phi_n(y-y_c)\phi_n(y^\prime-y_c)~,
\end{eqnarray}
and
\begin{eqnarray}\label{Grn21}
G_{21}({\bf r},{\bf r}^\prime,t) &=& \frac{1}{2\pi}\int_{-\infty}^{+\infty}dq_xe^{iq_x(x-x^\prime)}\\
&\times&\sum_{n=0}^{\infty}Q_{n+1}\phi_{n+1}(y-y_c)\phi_n(y^\prime-y_c)~, \nonumber
\end{eqnarray}
where, $P_n = \exp(-i {\gamma_nt})+  2 i A_n^{-1}\sin({\gamma_nt})$,  
$Q_n =  2 i D_n~A_n^{-1}~\sin({\gamma_nt})$ and
$\hbar \gamma_n=\sqrt{\Delta^2+\varepsilon^2n}$~.

Now inserting Eqs.~(\ref{Grn11})-(\ref{Grn21}) into Eq. (\ref{wavp_t}) it is straightforward to 
obtain the components of the wave packet at a later time $t$ in the following 
two component form:
\begin{eqnarray}
\begin{pmatrix}
 \Psi_1({\bf r},t)\\ \Psi_2({\bf r},t)
\end{pmatrix}
&=&\frac{1}{\sqrt{2}\pi l_c}\sum_{n=0}^\infty \frac{(-1)^n}{2^nn!N_n}
\int du e^{F(x,u)}u^n\nonumber\\
&\times&\begin{pmatrix}
P_{n+1}(t)\phi_n(y-y_c)\\Q_{n+1}(t)\phi_{n+1}(y-y_c)
\end{pmatrix},
\end{eqnarray}
where $F(x,u)=iux/l_c-(l_cq_{0}-u)^2/2-u^2/4$ with 
$u = q_x l_c$~.
The expectation values of the components of the position operator  can now be calculated from the following equation,
\begin{eqnarray}
\begin{pmatrix}
\langle x(t)\rangle\\ \langle y(t)\rangle
\end{pmatrix}
=\sum_{i=1}^2\int dxdy\Psi_{i}^\ast({\bf r},t)
\begin{pmatrix}
 x\\y
\end{pmatrix}
\Psi_i({\bf r},t).
\end{eqnarray}
The time dependent expectation value of the $x$ component of the position operator, 
can be split into two terms, $\langle x(t)\rangle=\langle x_1(t)\rangle+\langle x_2(t)\rangle$ with 
$\langle x_i(t)\rangle=\int d{\bf r}\Psi_i^\ast({\bf r},t)x\Psi_i({\bf r},t)$, and $i=1,2$. Now 

\begin{eqnarray}
\langle x_1(t)\rangle&=&\frac{1}{2\pi^2l_c^2}\sum_{n,n^\prime}
\frac{(-1)^{n+n^\prime}P^\ast_{n^\prime+1}P_{n+1}}{2^{n+n^\prime}n^\prime!n!N_{n^\prime}N_n}
\int dxdydudu^\prime \nonumber\\
&\times&e^{F(x,u)+F^\ast(x,u^\prime)}x\phi_n(y-y_c)\phi_{n^\prime}(y-y_c^\prime)~,
\end{eqnarray}
where $y_c^\prime=q_x^\prime l_c^2=l_cu^\prime$. 
Note that 
$\int dx ~x e^{i(u-u^\prime)x/l_c}=2\pi i~l_c^2~[\partial_{u^\prime}\delta(u^\prime-u)]$.
Using the following relation 
\be
\int du^\prime g(u^\prime) \left[\partial_{u^\prime}\delta(u^\prime-u)\right]
=- \left.\frac{dg(u^\prime)}{du^\prime} \right|_{u^\prime=u}~,
\ee
and doing a lengthy but straightforward calculation we  obtain
\begin{eqnarray} \label{eq:xint}
\langle x_1(t)\rangle&=&\frac{1}{i \pi }e^{-\frac{\tilde{q}_0^2}{3}}
\sum_{n,n^\prime}
\frac{(-1)^{n+n^\prime}P^\ast_{n^\prime+1}P_{n+1}}{2^{n+n^\prime}n^\prime!n!N_{n^\prime}N_n}
\int dydu~u^{n+n^\prime}\nonumber\\
&\times&e^{-(\sqrt{\frac{3}{2}}u-\sqrt{\frac{2}{3}}\tilde{q}_0)^2}
\Big\{\Big(\frac{n^\prime}{u}+\tilde{q}_0+\frac{y}{l_c}-\frac{5u}{2}\Big)\phi_{n^\prime}\nonumber\\
&-&\sqrt{2n^\prime}\phi_{n^\prime-1}(y-y_c)\Big\}\phi_n(y-y_c)~,
\end{eqnarray}
where $\tilde{q}_0=q_{0}l_c$. The integration over $y$ in Eq.~\eqref{eq:xint} can be done using the orthonormality of Hermite 
polynomials. Using the result: $\int_\infty^\infty dv~v^l~e^{-(v-\alpha)^2}
=\sqrt{\pi}~H_l(i\alpha)/(2i)^l$, and the recurrence relations for the Hermite 
polynomials we finally obtain
\be \label{eq:x1}
\langle x_1(t)\rangle=\sum_{n=0}^\infty  \xi_n ~\Im\left[P_{n+2}^\ast(t)P_{n+1}(t)\right]~,
\ee
where we have defined the time independent term,
\be
\xi_n = \frac{il_c}{3}e^{-\frac{\tilde{q}_0^2}{3}} \Big(\frac{-1}{12}\Big)^n \frac{1}{n!} H_{2n+1}\left(i\sqrt{\frac{2}{3}}\tilde{q}_0\right)~.
\ee
Following the same procedure as above,  we find 
\be \label{eq:x2}
\langle x_2(t)\rangle= \sum_{n=0}^\infty \xi_n ~\sqrt{\frac{n+2}{n+1}}~\Im\left[Q_{n+2}^\ast(t)Q_{n+1}(t)\right]~.
\ee

\begin{figure}[t]
\begin{center} 
\includegraphics[width=1.0 \linewidth]{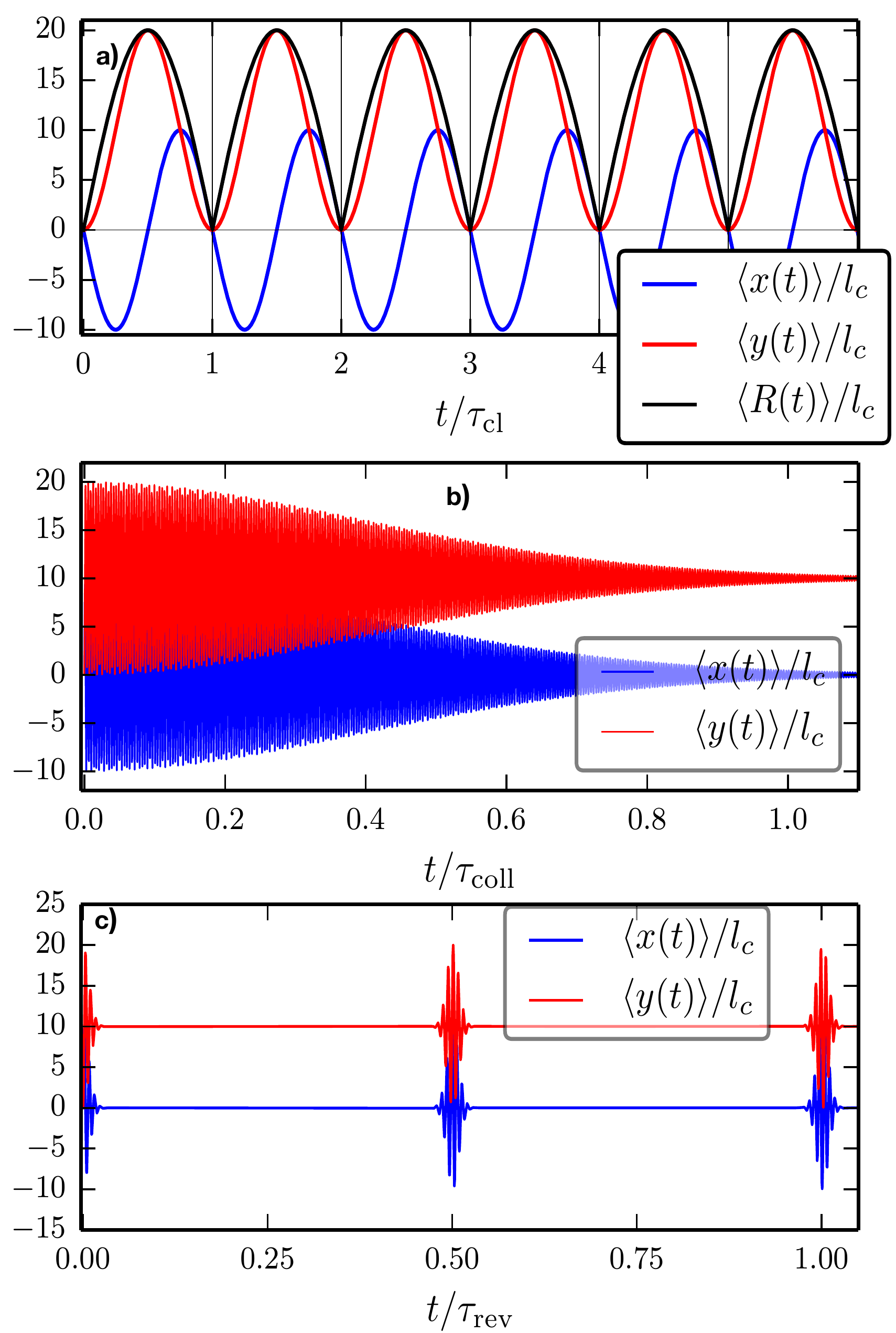}
\end{center}
\caption {
Panel a) shows the expectation value of $x$, $y$ and $R$ versus $t$ for a given width of the initial Gaussian wave packet, over a time period of a few $\tau_{\rm cl}$. Note that $\la x(t) \ra$ is centered around $0$, while $\la y \ra$ is centered around $y_c = q_0 l_c^2$, which is chosen to be $10$ (in units of $l_c$) for this plot. As a consequence, we see in Panel a) that $\la R(t) \ra \equiv \sqrt{ \la x (t) \ra^2 + \la y(t)\ra^2}$ closely follows $\la y\ra$ and this behaviour is maintained for larger timescales too --- {\it i.e.}, in Panels b) and c).
In Panel b) we plot the expectation value of $ x$ and $y$ versus $t$ over the collapse timescale,  $\tau_{\rm coll}$, to highlight the dephasing  of the wave packet amplitude due to quantum interference. Panel c) displays expectation value of $ x$ and $y$ versus $t$ over timescales of $\tau_{\rm rev}$, and highlights the spontaneous collapse and revival of the amplitude of the incident wave packet. Other parameters are chosen to be $\Delta = 0.8$ eV, $v_F = 85000$ m/s, with $l_c = 2\times 10^{-8}$ m, and $q_0 l_c = 10$ for these plots. }
\label{fig2}
\end{figure}

The calculations for the expectation value of the $y$ component of the position 
operator are less involved, and we obtain the following expression for 
$\langle y(t)\rangle=\langle y_1(t)\rangle+\langle y_2(t)\rangle$ with, 
\be \label{eq:y1}
\langle y_1(t)\rangle=\sum_{n=0}^\infty \xi_n ~\Big(\Re\left[P_{n+2}^\ast P_{n+1}\right]- \left|P_{n+1} \right|^2\Big)~,
\ee
and
\be \label{eq:y2}
\langle y_2(t)\rangle= \sum_{n=0}^\infty \xi_n \left(\sqrt{\frac{n+2}{n+1}} ~ \Re\left[Q_{n+2}^\ast Q_{n+1}\right]-\left| Q_{n+1}\right|^2\right)~.
\ee
Note that in the above equation we have suppressed the time arguments. 

We plot the expectation values of the position operator, given by Eqs.~\eqref{eq:x1}-\eqref{eq:y2} in Fig.~\ref{fig2} over various timescales. It is useful to have an estimate of the various timescales discussed in the previous subsection for our present case.
For monolayer MoS$_2$, and the choice of the initial wave packet in Eq.~\eqref{ini_mag}, it is easy to show that $n_0 \approx l_c^2 q_0^2/2$ (for example, see Appendix A of Ref.~[\onlinecite{Kramer}]). Now, using the explicit form of the Landau level energies given by Eq.~\eqref{Enr} in Eq.~\eqref{timescale1}, we obtain the timescales for our case to be 
\be
\tau_{\rm osc} = \frac{2 \pi \hbar}{\Omega_0} ~,~~~ \tau_{\rm cl} = \frac{4 \pi \hbar \Omega_0}{\varepsilon^2}~, ~{\rm and}~~~ \tau_{\rm rev} = \frac{16 \pi \hbar \Omega_0^3}{\varepsilon^4}~,
\ee
where we have defined $\Omega_0  \equiv \sqrt{\Delta^2 + \varepsilon^2 n_0}$. Note that for realistic values for MoS$_2$ parameters --- see caption in Fig.~\ref{fig2},  these timescales are: $\tau_{\rm osc} = 5.10$ fs, $\tau_{\rm cl} = 428 $ ps, $\tau_{\rm rev} = 71.8~ \mu$s, and finally $\tau_{\rm coll} = T_{\rm rev}/n_0 = 1.43 ~\mu$s. Here we have used the fact the for a coherent state wave packet --- see Eq.~\eqref{ini_mag}, $\delta n = \sqrt{n_0}$.

It is noteworthy that the dominant contribution to the sum in Eqs.~\eqref{eq:x1}-\eqref{eq:y2}, arises from the neighbourhood of $n \approx n_0 $. Now for $ (\varepsilon/\Delta)^2 n \ll 1$, $ \hbar\gamma_n \approx \Delta + \varepsilon^2 n/(2 \Delta) $, and it can be shown that Eqs.~\eqref{eq:x1}-\eqref{eq:y2}, describe the quasiclassical cyclotron motion given by 
\bearr \label{xcl}
\la x(t) \ra  &=&  \sqrt{2 n_0} ~l_c ~\sin\left(\frac{2 \pi t}{\tau_{\rm cl}} \right)~, \\ \label{ycl}
\la y(t) \ra  &=&  \sqrt{2 n_0} ~l_c ~\left[1- \cos\left(\frac{2 \pi t}{\tau_{\rm cl}} \right)\right]~. 
\eearr
This is also clearly evident from Fig.~\ref{fig2}a). Note however, that this quasiclassical description is only valid for timescales of the order of $\tau_{\rm cl}$, as we have ignored all the terms which lead to quantum interference and consequently lead to the wave packet amplitude collapse and revival.

\begin{figure}[t]
\begin{center} 
\includegraphics[width=1.0 \linewidth]{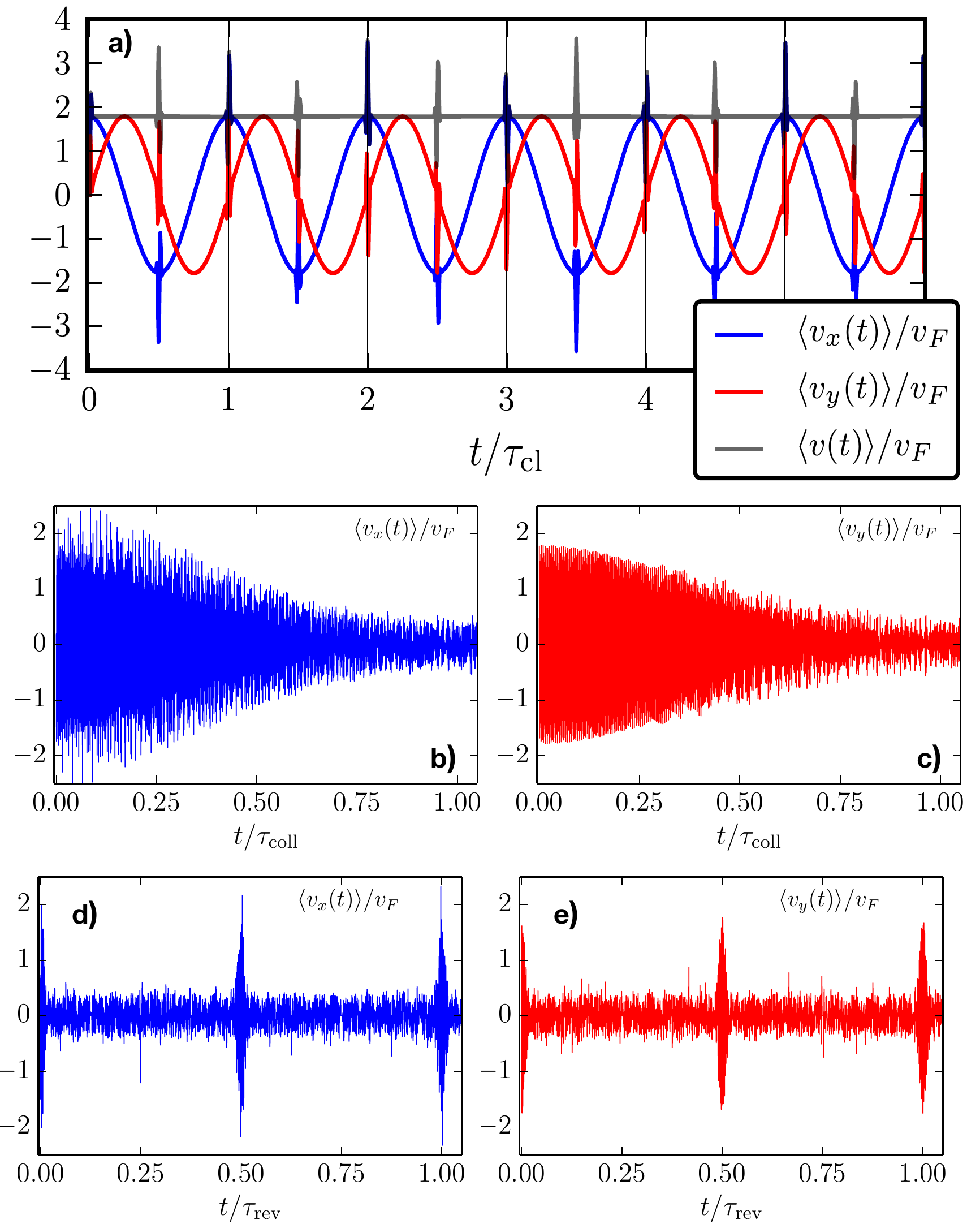}
\end{center}
\caption{Expectation value of the velocity operator over various timescales. Panel a) shows the expectation value of $v_x$, $v_y$ and $\la v \ra \equiv \sqrt{ \la v_x\ra^2 + \la v_y\ra^2}$ versus $t$ for a given width of the initial Gaussian wave packet, over a time period of a few $\tau_{\rm cl}$. In Panel b) we study the expectation value of $ v_x$ and $v_y$ versus $t$ over the collapse timescale $\tau_{\rm coll}$, to highlight the dephasing in the velocity expectation value. Panel c) displays expectation value of $ v_x$ and $v_y$ versus $t$ over timescales of $\tau_{\rm rev}$, and highlights the spontaneous collapse and revival of the velocity due to quantum interference terms arising from higher derivatives of the discrete Landau level spectrum. Other parameters are same as in Fig.~\ref{fig2}. 
}
\label{fig3}
\end{figure}

To calculate the expectation value of the velocity operator, we use the Heisenberg equation of motion: $\hbar \hat{v}_j = i [\hat{H}, {\hat r}_j]$, and  straightforward calculations yield the following expressions for the expectation value for the velocity operator:
\be \label{velB1}
\langle v_x(t)\rangle = \Re [\la v(t) \ra]~,~~{\rm and},~~\la v_y(t)\ra = \Im [\la v (t)\ra]~,
\ee
where 
\be \label{velB2}
\la v(t) \ra = \frac{\sqrt{2} v_F}{l_c} \sum_{n=0}^\infty \xi_n \frac{1}{\sqrt{1+n}} ~P_{n+2}^\ast Q_{n+1}~.
\ee
As in the case of the expectation value of the position operator, for timescales of the order of $\tau_{\rm cl}$, it can be shown that Eqs.~\eqref{velB1}-\eqref{velB2}
reduce to the quasiclassical cyclotron velocities given by 
\bearr \label{vxcl}
\la v_x(t) \ra  &=&  \sqrt{2 n_0} ~l_c~\frac{2 \pi}{\tau_{\rm cl}} ~\cos\left(\frac{2 \pi t}{\tau_{\rm cl}} \right)~, \\ \label{vycl}
\la v_y(t) \ra  &=&  \sqrt{2 n_0} ~l_c~\frac{2 \pi}{\tau_{\rm cl}} ~ \sin\left(\frac{2 \pi t}{\tau_{\rm cl}} \right)~. 
\eearr
Similar expressions for velocity in the quasiclassical regime, were also obtained in Ref.~[\onlinecite{Demikhovskii_PRA}], where they studied the dynamics of relativistic wave packets.
Note that these can also be obtained by differentiating Eq.~\eqref{xcl} and Eq.~\eqref{ycl} respectively, and are also consistent with Fig.~\ref{fig3}a).
We also note that in the $\Delta \to 0$ limit, the MoS$_2$ Landau level spectrum reduces to graphene Landau level spectrum and in the same limit, our result for velocity in Eqs.~\eqref{velB1} and \eqref{velB2} are similar to the known results for cyclotron dynamics in graphene in Ref. [\onlinecite{Romera_PRB}], where they had a different initial wave packet.  We plot the expectation value of the velocity operator over various timescales in Fig.~\ref{fig3}, and find a behaviour similar to that of the position operator, {\it i.e.}, quasiclassical cyclotron orbit oscillations over $\tau_{\rm cl}$, and then wave packet amplitude collapse and revival over $\tau_{\rm rev}/2$. 

\section{Summary}
In this Article, we have studied the dynamics of a Gaussian wave packet in monolayer MOS$_2$, both in the presence of a perpendicular magnetic field and without it. In the absence of a magnetic field, we obtain explicit expressions for the expectation values of the position and velocity operators, which show the expected behaviour of amplitude oscillations (ZB) whose period is given by Eq.~\eqref{eq:tauzb}, and which decay over timescale given by Eq.~\eqref{eq:decay}. 

For a system with discrete Landau level spectrum ({\it i.e.}, in the presence of a perpendicular magnetic field), injecting an initial electron wave packet which is peaked around some Landau level energy of MoS$_2$, we find that the wave packet initially evolves quasiclassically and oscillate with a period of $\tau_{\rm cl}$ (the cyclotron time-period). However at later times, the wave packet eventually spreads and quantum interference leads to its `collapse', and at even longer times, 
that are multiples (or rational fractions) of $\tau_{\rm rev}$, the wave packet is revived, and the electron position and velocity regains its initial amplitude --- again undergoing quasiclassical oscillatory motion. At much shorter timescales, {\it i.e.,} $\tau_{\rm osc} \ll \tau_{\rm cl} \ll \tau_{\rm rev}$, the electron wave packet undergoes {\it zitterbewegung}, but this amplitude is much smaller than that of the quasiclassical cyclotron motion, and is thus suppressed.

We emphasize here that for typical parameters of MoS$_2$, $\tau_{\rm osc} = 5.10$ fs, $\tau_{\rm cl} = 428 $ ps, and finally $\tau_{\rm rev} = 71.8~ \mu$s, and as a consequence, wave packet revivals manifested in velocity (and consequently in current) is more accessible to experimental probing. In addition, such dynamics can also be investigated by experimentally studying the nature of electromagnetic multipole radiation and absorption. The results obtained in this Article, can also be useful for the observation of spontaneous collapse and revival in experiments with trapped ions \cite{Gerritsma}.

\section{Acknowledgements}  A.A. gratefully acknowledges funding from the INSPIRE faculty fellowship by DST  (Govt. of India), and from the Faculty Initiation Grant by IIT Kanpur, India.

\end{document}